
\font\fontbig=cmr10 scaled \magstep2

\def\L{{\cal L}}
\def\del{\partial}

\mathchardef\lag="724C
\magnification=1200
\hsize=16.0 true cm
\baselineskip 15pt

\def\pl{{\sl Phys.\ Lett.\ }}
\def\np{{\sl Nucl.\ Phys.\ }}
\def\pr{{\sl Phys.\ Rev.\ }}
\def\prl{{\sl Phys.\ Rev.\ Lett.\ }}

\rightline{BIHEP-TH-95-23}
\rightline{December 1995}
${}^{}$
\vskip 4pc
\centerline{\bf {\fontbig FERMION NUMBER VIOLATION IN HEAVY }}
\centerline{\bf {\fontbig FERMION DECAY }}
\vskip 4pc
\centerline{\bf Keyan Yang }
\vskip 1pc
{\sl
\centerline{\bf Institute of High Energy Physics, Academia Sinica}
\centerline{\bf Beijing~100039, China}
}
\vskip 4pc
\centerline{\bf ABSTRACT }
\vskip 2pc
The presence of a heavy fermion doublet of fourth
generation in the standard model would lead to a anomalous decay
with fermion number non-conservation.
The anomalous decay path in the background of the electroweak
instanton is demonstrated by numerical calculation.
If the mass of fermion exceeds the critical value
$m_f^{\rm cr}$=9.2 TeV (for $M_H=M_W$),
it is shown that there exists the rapid anomalous decay of the heavy
fermion in semi-classical approximation.
The dependence of the critical fermion mass on the Higgs mass is also
presented.
\vskip 2pc
\par
PACS number(s): 11.15.Kc,12.15.Ji
\vfil\eject
\centerline{\bf I. INTRODUCTION}
\vskip 3pc
Recent experimental observation for top quark with large mass around
176 GeV has been reported by CDF Collaboration[1].
The top quark mass is much bigger than the masses of other quarks.
There is a large mass hierarchy among three generation fermions.
The ratio of $m_u$:$m_c$:$m_t$ is approximately as 1:260:35200.
On the other hand, present experimental accuracy is not
sufficient to exclude the existence of one extra generation of leptons
and quarks[2]. The new fermions are supposed to possess the same colour
and electroweak quantum numbers as the ordinary ones and to mix very
tinily with the ordinary three generation.
If the fourth generation quarks exist, it would be very heavy
according to the large mass ratio for up, charm and top.
The mass of fourth generation quarks would be over 10 TeV because
the mass ratio between two generations is always over 100 times.
So, it is very interesting to investigate the physics of standard model
with very heavy fermion.
\par
When the heavy fourth generation quarks represents in the standard model,
one of the most important effects is that the non-perturbative electroweak
fermion number non-conservation may be naturally unsupressed[3].
In this paper we consider the anomalous decay of the heavy fourth generation
quarks in standard electroweak model.
We are known that the baryon number and lepton number do not absolutely
conserve in the standard model due to the quantum anomaly[4].
The process for baryon number and lepton number nonconversation is
spontaneous fermion number violation due to instanton induced transitions
between topologically distinct vacua[5].
Under normal conditions, the amplitudes with nonconserved baryon number
that are due to this mechanism are suppressed by the factors
${\rm exp}(-8\pi^2/g_W^2)\sim 10^{-170}$, where
$g_W^2 =e^2 /\sin^2 \theta_W$ is the coupling constant of the electroweak
gauge group $SU(2)_L$. However,
it has recently been realized that there can be a
great amplification of anomalous fermion number nonconservation[6].
Generally, this might occur when the energy stored in the system is big
enough. In principle, the energy can be of different forms.
The simplest condition is provided by the system at high-temperatures[7],
or collisions of particles at high-energies[8], and in decays of heavy
particles[9]. The characteristic energy scale, at which the anomalous
process become rapid, is set by the sphaleron mass, which determines the
height of the energy barrier between the topologically distinct vacua,
and is of order 10 TeV.
\par
The mechanism of fermion number nonconservation can be directly shown
in the fermion level crossing picture[10,11].
When the bosonic field configuration evolves from the trivial vacuum
configuration to the topologically distinct one with unit topological
number, the fermion energy level emerges from the positive continuum,
crosses zero and disappears in the negative continuum. At small fermion
masses, the state corresponding to the fermion is separated from
the topologically non-trivial vacuum state by a finite energy barrier.
The fermion can decay into the vacuum state due to semi-classical
tunneling through this barrier, but the probability of such a process is
exponentially small. If the mass of the fermion is comparable with the
height of the energy barrier, the fermion may become classically
unstable with respect to the anomalous decay. In this case the barrier
between the fermion and the topologically non-trivial vacuum
disappears, the decay, instead of tunneling, proceeds classically and
the exponential suppression of the amplitude is absent.
\par
Anomalous fermion number non-conservation in decays of the system of
elementary fermions has been investigated by Rubakov[3] at first,
It was shown that the mass of fermion exceeds some critical values,
the barrier disappears and the system freely rolls down, decaying
into the vacuum state with zero fermion number. Recently, Petriashvili
[9] has calculated the critical values of fermion mass
by using the variational ansatz.
In this paper, we will demonstrate the whole path of anomalous
decay of heavy fermion in an electroweak instanton by numerical
calculation and give the critical values of fermion mass for unsuppressed
fermion number non-conservation appearing.
For simplicity, we assume that the charged and neutral components of
the fourth generation doublet have equal masses and neglect the weak
hypercharge interactions $U(1)$.
Throughout this paper we work in the classical approximation, i.e.
we neglect the radiative corrections due to the boson loops and the
contribution of the Dirac sea to the energy of the system.
For strong Yukawa coupling, this approximation is far from
being convicing, nevertheless, we do hope that our results are
qualitatively correct.
\par
On the other hand, the heavy fermion in the standard model cannot be
included without violating vacuum stability[12]. However,
the problem may be overcome by introducing new physics in the TeV region,
for example, the arguments of ref. 12 are not valid
if there exist heavy scalars with roughly the same masses,
as can be the case in supersymmetric theories.
Therefore, we think that the decays of heavy fourth generation quarks
can be of phenomenological interest.
\par
We work in the $A_0(x)=0$ gauge and focus upon the Euclidean time
parameter $x_0$. As $x_0$ changes from $-\infty$ to
$+\infty$, the one-instanton field evolves from one pure gauge
configuration to another topologically distinct pure gauge.
The three-dimensional Dirac Hamiltonian in the presence of such a field
depends parametrically on $x_0$
through $A_i(x_0,{\bf x})$. The spectrum of eigenenergy of fermion as a
function of $x_0$, gives information about the behavior of the quantized
Dirac field in the adiabatic approximation. Then, we can calculate the
total energy of the system which depend on $x_0$. If the total
energy monotonically decreases (energy barrier is absent), one can expect
that rapid anomalous decay of fermion might occur classically.
\par
The remainder of this paper is organized as follows.
Section II introduce the Weinberg-Salam model Lagrangian with two
approximations employed and the formulas of fermion number violation
in electroweak theory. The electroweak instantion solution and its
expression in the Minkowski space are given in Sec. III. In this section,
the Chern-Simons number and the static energy of the instanton are presented
as a function of Euclidean time. In Sec. IV we derive the radial equations
for fermions. We present the anomalous decay path of the heavy fermion
in the background of instantons by numerical calculation in Sec. V.
Finally, in Sec. VI, a brief discussion is given.

\vskip 3pc

\centerline{\bf II. FERMION NUMBER VIOLATION }
\vskip 2pc
Let us consider the bosonic sector of the Weinberg-Salam model in the limit
of vanishing mixing angle. In this limit the $U(1)$ field
decouples and can consistently be set to zero:
\vskip 1 pt
$$
    \L_b =-{1 \over 4 }F_{\mu \nu}^a F^{\mu \nu,a} + (D_\mu \Phi)^\dagger
          (D^\mu \Phi) - \lambda (\Phi^\dagger \Phi - {v^2 \over 2} )^2
          \eqno (2.1)
$$
with the $SU(2)_L$ field strength tensor,
$$
    F_{\mu\nu}^a =\del_\mu A_\nu^a - \del_\nu A_\mu^a + g\epsilon^{abc}
                  A_\mu^b A_\nu^c  ,    \eqno (2.2)
$$
and the covariant derivative for the Higgs field,
$$
     D_\mu \Phi = (\del_\mu -{i \over 2}g\tau^a A_\mu^a)\Phi , \eqno (2.3)
$$
where $g$ is the gauge coupling constant and in electroweak theory we employ
the value $g=0.67$. $A_\mu^a (x)$ $(a=1,2,3,)$ are real vector fields and can
be described as a matrix field $A_\mu (x)={1 \over 2}g\tau^a A_\mu^a (x)$,
$\tau^a$ being the isospin Pauli matrices.
\par
The $SU(2)_L$ gauge symmetry is spontaneously broken due to the non-vanishing
vacuum expectation value $v$ of the Higgs field,
$$
      \langle \Phi \rangle ={v \over \sqrt2}{0\choose 1}   \eqno (2.4)
$$
leading to the boson masses
$$
      M_W=M_Z={1 \over 2}gv, \qquad M_H=v\sqrt {2\lambda}, \eqno (2.5)
$$
where we take $v$=246GeV.
\par
The model has a non-trivial vacuum structure.
The classical vacuum configuration, being defined by the minima of the
bosonic part of static energy
\vskip 1pt
$$
       E_b=\int d^3x\big\{-{1 \over 4 }F_{\mu \nu}^a F^{\mu \nu,a}
           + (D_\mu \Phi)^\dagger
          (D^\mu \Phi) + \lambda (\Phi^\dagger \Phi - {v^2 \over 2} )^2
          \big\}
         \eqno (2.6)
$$
are represented by the pure gauge configurations
$$
         A_\mu(x)=U(x)\partial_\mu U^{-1}(x), \quad \quad
          \Phi(x)=U(x)\phi_0,  \eqno (2.7)
$$
where $U(x)$ is a $2\times 2$ unitary matrix of SU(2); $\phi_0$
is a constant isospinor and takes the value of the Higgs field for
$A_\mu=0$. The classical vacua have a discrete set labeled by the
integer $q$, and one vacuum differs from another topologically by
nontrivial gauge transformations which carry an topological number
$$
   q={g^2 \over 16\pi^2}\int d^4x\epsilon^{\mu\nu\rho\sigma}
      F_{\mu\nu}^a F_{\rho\sigma}^a .  \eqno   (2.8)
$$
The topological current is
$$
     K^\mu = {g^2 \over 8\pi^2}\epsilon^{\mu\nu\rho\sigma}{\rm Tr}
             (A_\nu \del_\rho A_\sigma -i{2 \over 3}gA_\nu A_\rho A_\sigma) .
             \eqno (2.9)
$$
All gauge field configurations can be classified by the Chern-Simons number
given by
$$
            N_{CS} =\int d^3 x K^0 .   \eqno  (2.10)
$$
The Chern-Simons number $N_{CS}$ may be regarded as a coordinate in
gauge-orbit space which measures the position of topologically
inequivalent vacua. For the vacua the Chern-Simons number is identical to
the integer, for nonvacuum it may take on arbitrary real values.
Neighbouring vacua with different topological numbers are separated by
a finite energy barrier. The height of the barrier is determined by the
sphaleron solution of the bosonic field equations[13]. This saddle-point
solution has Chern-Simons number $N_{CS}=1/2$, and its energy is equal
to the height of the barrier.
\par
The most straightforward realization of a fermionic extension of the
standard model is the introduction of a fourth generation of fermions.
The new fermions are supposed to possess the same colour and electroweak
quantum numbers as the ordinary ones.
For vanishing mixing angle, considering the heavy fermion doublet degenerate
in mass, the heavy fermion Lagrangian in the chiral representation reads,
\vskip 1 pt
$$
  \eqalign{  \L_f = &{\bar q}_L i\gamma^\mu D_\mu q_L+{\bar u}_R
                     i\gamma^\mu \del_\mu u_R
                   + {\bar d}_R i\gamma^\mu \del_\mu d_R       \qquad  \cr
                 &-f_q {\bar q}_L ({\tilde \Phi} u_R +\Phi d_R)
                  -f_q ({\bar d}_R \Phi^\dagger
                  +{\bar u}_R {\tilde \Phi}^\dagger )q_L,
}     \eqno (2.11)
$$
where $q_L$ denotes the lefthanded doublet $(u_L, d_L)$, $u_R$ and $d_R$ are
the righthanded singlets, with covariant derivative,
$$
    D_\mu q_L = (\del_\mu - {i \over 2}g\tau^a A_\mu^a )q_L , \eqno (2.12)
$$
and with $\tilde\Phi=i\tau_2 \Phi^{*}$. The heavy fermion mass is given by
$$
      m_u^{}=m_d^{} = m_f^{} = {1 \over \sqrt2} f_q v .  \eqno (2.13)
$$
The gauge-invariant current of the doublet $J^\mu ={\bar q}_L\gamma^\mu q_L$
is conserved at the classical level, but is anomalous at the quantum level:
$$
    \del_\mu J^\mu ={g^2 \over 16\pi^2}\epsilon^{\mu\nu\rho\sigma}
                    F_{\mu \nu}^a F_{\rho\sigma}^a .    \eqno  (2.14)
$$
\par
The integration of right-hand side of the equation (2.14) is expression of
the topological charge of a gauge field configuration as (2.8).
The equation (2.14) indicates that the number of fermions may not be
conserved, the changes of baryon number $B$ and lepton number $L$ are
given as
$$
       \Delta B =\Delta L =n_f^{} q  ,      \eqno (2.15)
$$
where $n_f^{}$ is the number of generations.

\vskip 3pc

\centerline{\bf III. EXPRESSION OF INSTANTON IN MINKOWSKI SPACE}
\vskip 2pc

The fermion number non-conservation in the electroweak theory is associated
with instanton which takes Euclidean field configuration[14].
For $SU(2)$ gauge field, an explicit solutions
with topological charge $q = 1$ in the regular gauge is given by
$$
\eqalign{
  A_0 (x) &=-{{i{\bf \tau} \cdot {\bf x}} \over {x^2 +\rho^2 }} ,  \cr
  {\bf A}(x)&=-{{i({\bf \tau} x_0 +{\bf \tau} \times {\bf x})}
                \over {x^2 +\rho^2 }} ,
}  \eqno  (3.1)
$$
where $x^2 =x_0^2 +{\bf x}^2$ and $\rho$ is some arbitrary scale parameter,
often referred to as the instanton size. The instanton can be viewed as a
solution of the Euclidean gauge field equations in which a vacuum at
$x_0 =-\infty$ evolves by propagation in imaginary time to a different
vacuum at $x_0 =+\infty$.
\par
In the Weinberg-Salam model, due to the presence of the Higgs field,
strictly speaking, for $v \not= 0$, there does not exist a finite action
solution of the classical Euclidean equations of motion.
However, as long as the size of instanton $\rho$ is not too large compared
to the inverse size of the order parameter $v$ in the Weinberg-Salam model,
as $\rho v\ll 1$, one can find a good approximate solution in the
Weinberg-Salam model[15]. This approximate solution
still has the gauge field configurations given by the $SU(2)$ instanton of
(3.1), while the Higgs doublet field takes the form,
$$
    \Phi = {{x_0 -i{\bf \tau}\cdot {\bf x}} \over \sqrt{x^2 +\rho^2}}
             {v \over \sqrt2 }\Phi_0   \eqno  (3.2)
$$
where $\Phi_0$ is a constant $SU(2)$ spinor and we can take
$\Phi_0 ={0 \choose 1}$ and $\tilde\Phi_0 ={1 \choose 0}$.
\par
In the Minkowski space version for instanton, it describes the tunneling
process in potential barrier which connects one vacuum to another
vacuum. In the gauge $A_0=0$, the instanton determines the path in the
configuration space connecting the vacua with different topologcal
number, $x_0$ being the parameter along the path. If the starts from
the trivial gauge vacuum ($N_{CS}$=0) with filled Dirac sea plus one real
fermion, and evolves to the non-trivial gauge vacuum with $N_{CS}=1$,
then the final state has filled Dirac sea and no real fermion,
so that the fermion number is not conserved.
\par
To cast the vector potential $A_\mu (x)$ given in (3.1) to the form of
the $A_0 (x)=0$, we make a gauge transformation $V(x)$ on $A_\mu (x)$
into the temporal gauge:
$$
      A'_0 (x)=V^{-1}(x)A_0 (x)V(x)+V^{-1}(x)\del_0 V(x) = 0   \eqno  (3.3)
$$
and define
$$
   {\bf A}'(x)=V^{-1} (x){\bf A}(x)V(x) +V^{-1}(x){\bf \nabla} V(x).
              \eqno  (3.4)
$$
Under this gauge transformations, the Higgs field $\Phi$ can be transformed
accordingly as
$$
      \Phi' (x)=V^{-1} (x)\Phi (x) .  \eqno  (3.5)
$$
{}From equation (3.3), we can find the solution for $V (x)$ by
integration as
$$
      V(x) =\exp [i{\bf \tau}\cdot \hat{\bf x} f(x)]  \eqno  (3.6)
$$
with
$$
     f(x) ={\sqrt{{\bf x}^2} \over \sqrt{{\bf x}^2 +\rho^2}} \big [\tan^{-1}
           ({x_0 \over \sqrt {{\bf x}^2 +\rho^2}}) +\theta({\bf x})\big ],
            \eqno  (3.7)
$$
where $\theta({\bf x})$ is a time-independent residual gauge freedom with
respect to the $A_0 (x)= 0$ gauge. We can  choose it to be constants,
$$
      \theta = (n+{1 \over 2})\pi , \qquad n=0,1,2, \cdots   \eqno  (3.8)
$$
\par
By substituting (3.1) and (3.2) into (3.4) and (3.5), respectively,
we obtain the general spherically symmetric form for the gauge field
${\bf A}' (x)$ and Higgs field $\Phi' (x)$ as
$$
\eqalign{
 {\bf A}'(x) &={1 \over g}\big \{ a(r)({\bf \tau} \times \hat{\bf x})
      +b(r)\big [ {\bf \tau} -({\bf \tau}\cdot \hat{\bf x})\hat {\bf x}\big ]
      +c(r)({\bf \tau}\cdot \hat {\bf x})\hat {\bf x}\big \}, \cr
  \Phi'(x) &={v \over \sqrt2 }\big [ h(r)+i{\bf \tau} \cdot \hat {\bf x}k(r)
              \big ]\Phi_0
}   \eqno  (3.9)
$$
with
$$
\eqalign{a(r)&=-{1 \over x^2+\rho^2}(r\cos 2f+x_0\sin 2f)-{\sin^2 f\over r},
                          \cr
 b(r)&=-{1 \over x^2+\rho^2}(x_0\cos 2f -r\sin 2f) -{\sin 2f \over 2r}, \cr
 c(r)&=-{x_0 \over x^2+\rho^2} -{df \over dr},  \cr
 h(r)&={1 \over \sqrt {x^2+\rho^2}}(-r\sin f+x_0\cos f), \cr
 k(r)&={1 \over \sqrt {x^2+\rho^2}}(-r\cos f-x_0\sin f),
 }     \eqno  (3.10)
$$
where $x^2=x_0^2+r^2$, $r =\sqrt {{\bf x}^2}$ and $\hat{\bf x}$ is a unit
three-vector in the radial direction given by
$\hat{\bf x} = {\bf x} /r$.
\par
By using (3.9), the Chern-Simons number $N_{CS}$ in (2.10) can be given by
$$
     N_{CS}(x_0) ={2 \over \pi}\int_0^\infty r^2dr\big [ 2c(a^2 +b^2
                  +{1 \over r}a)+(ba'-ab')\big ]    \eqno  (3.11)
$$
where the prime means differentiation with respect to $r$.
Due to $x_0$ dependence of the functions $a(r)$, $b(r)$ and $c(r)$,
the $N_{CS}$ is the function of $x_0$. We do not now
attribute any physical significance to the variable $x_0$ and regard
${\bf A}({\bf x},x_0)$ and $\Phi ({\bf x},x_0)$ as a path in the
configuration space, $x_0$ being simply a parameter along this path.
In this way, $x_0$ can describe the configuration space path as same as
the $N_{CS}$. $N_{CS}$ changes from 0 to 1 when $x_0$ varies
from $-\infty$ to $+\infty$.
\par
The static energy of gauge and Higgs fields for the configurations
at some fixed value of the parameter $x_0$ can be obtained by using (3.9):
$$
\eqalign{
  E(x_0)=& {4\pi \over g^2}\int_0^\infty dr\biggr \{
            2r^2(a^2+b^2+{1 \over r}a)^2   \cr
     &+r^2(a' +{1 \over r}a +2bc)^2
      +r^2[b'+{1 \over r}b-{1 \over r}(1+2ra)c]^2   \cr
     &+g^2v^2\big\{ (k^2+h^2)[1+2ra+r^2(2a^2+2b^2+c^2)]  \cr
     &+(1+2ra)(k^2-h^2)-4rbhk +r^2(h'^2+k'^2)  \cr
     &-2r^2c(k'h-kh') +v^2\lambda r^2 (h^2+k^2-1)^2 \big\} \biggr \}
}      \eqno   (3.12)
$$
where the prime means differentiation with respect to $r$.
$E$ is the function of $x_0$. We can define parametrically the energy $E$ as
a function of $N_{CS}$.
\par
By use of the functions in (3.10), one can calculate the static
energy $E$ for varying $x_0$. There exists a symmetric potential barrier,
while $E(x_0=\pm \infty) =0$ since the gauge and Higgs field tend to vacuum
configuration as $x_0 \to \pm \infty$ and $E(x_0 =0)$ is at the top
of the barrier. The top of static energy $E(0)$ is given by
$$
         E(0)=\pi^2({3 \over g^2}{1 \over \rho}+
              {3 \over 8}v^2\rho+{1 \over 4}\lambda v^4\rho^3).  \eqno (3.13)
$$
The top energy E(0) takes the minimum value when the instanton size
$\rho$ takes
$$
         \rho_0={4 \over g\sqrt{1+\sqrt{1+8\alpha^2}}}{1 \over v}
                \eqno (3.14)
$$
where $\alpha$ is given by $\alpha=M_H/M_W$ and we take
$\lambda=g^2\alpha^2/8$. Substituting (3.14) into (3.13),
the minimum barrier height $E^{\rm min}(0)$ is given by
$$
      E^{\rm min}(0)={\pi^2 \over g}\Bigr[
                 {3 \over \sqrt{1+\sqrt{1+8\alpha^2}}}
                +{8\alpha^2 \over (1+\sqrt{1+8\alpha^2})^{3/2}}\Bigr] v.
                     \eqno (3.15)
$$
We show the minimum barrier in Fig. 1 where the instanton size takes
the value of (3.14) with $\alpha=1$.
In the following to calculate the critical mass
of the fermion at which the exponential suppression for fermion number
non-conservation disappears, we will use this minimum barrier for
our calculation.

\vskip 3pc

\centerline{\bf IV. THE RADIAL EQUATIONS FOR HEAVY FERMIONS}
\vskip 2pc
Let us now consider heavy fermions in the background fields of the
electroweak instanton with the form of (3.9).
To retain spherical symmetry we consider the heavy fermion doublets
degenerate in mass. From the fermion Lagrangian (2.11),
for each value of Euclidean time $x_0$, we obtain the time-independent
Dirac equations for the lefthanded doublet
\vskip 1 pt
$$
    iD_0 q_L +i\sigma^i D_i q_L -f_q (\tilde\Phi u_R +\Phi d_R) =0
    \eqno  (4.1)
$$
and for the righthanded singlets
\vskip 1 pt
$$
\eqalign{
  &i\del_0 u_R -i\sigma^i\del_i u_R -f_q {\tilde \Phi}^\dagger q_L =0 , \cr
  &i\del_0 d_R -i\sigma^i\del_i d_R -f_q  \Phi^\dagger q_L =0 ,
}       \eqno  (4.2)
$$
where $\sigma^i$ are Pauli spin matrices. Wavefunctions $q_L$ ,$u_R$ and
$d_R$ depend on $x_0$ which enters only as a parameter.
\par
Spherically symmetric fermion fields are described by the ansatz:
\vskip 1 pt
$$
\eqalign{  q_L(r) &=e^{-i\epsilon t}\big [G_L(r)+i{\bf \sigma}\cdot {\bf x}
          F_L(r)\big ]\chi_h^{} , \cr
 u_R(r) &=e^{-i\epsilon t}\big [G_R(r)+i{\bf \sigma}\cdot {\bf x} F_R(r)\big ]
                \chi_1^{} , \cr
 d_R(r) &=e^{-i\epsilon t}\big [G_R(r)+i{\bf \sigma}\cdot {\bf x} F_R(r)\big ]
                \chi_2^{}
}   \eqno  (4.3)
$$
with
\vskip 2 pt
$$
\eqalign{
     \chi_h^{} &={1 \over \sqrt 2}\biggr [ {1 \choose 0}_S{0 \choose 1}_I
             -{0 \choose 1}_S{1 \choose 0}_I \biggr ] ,  \cr
     \chi_1^{} &={1 \over \sqrt 2} {1 \choose 0}_S , \qquad
     \chi_2^{}  =-{1 \over \sqrt 2}{0 \choose 1}_S ,
}        \eqno  (4.4)
$$
where $S$ refers to spin , $I$ to isospin.
$\chi_h^{}$ is  the hedgehog spinor satisfying the spin-isospin relation
$\sigma\chi_h^{} + \tau\chi_h^{} =0$. $\chi_1^{}$, $\chi_2^{}$ are the
constant spin spinors which can construct the hedgehog spinor $\chi_h^{}$
with isospin spinors $\Phi_0$ and ${\tilde \Phi}_0$.
\par
By using these ansatz for equations (4.1) and (4.2), we obtain the
following set of four coupled first order differential equations:
\vskip 1 pt
$$
\eqalign{&F'_L+({2 \over \tilde r }+2\tilde a )F_L+(2\tilde b +\tilde c )G_L
           -\tilde\epsilon G_L=-(\tilde h G_R +\tilde k F_R), \cr
 &G'_L -2\tilde a G_L +(2\tilde b -\tilde c )F_L+\tilde\epsilon F_L
            =(\tilde h F_R -\tilde k G_R), \cr
 &F'_R +{2 \over \tilde r }F_R+\tilde\epsilon G_R
            =(\tilde h G_L -\tilde k F_L), \cr
 &G'_R-\tilde\epsilon F_R =-(\tilde h F_L +\tilde k G_L),
}  \eqno  (4.5)
$$
where $\tilde r =m_f r$, $\tilde\epsilon =\epsilon / m_f^{}$ and the
prime means differentiation with respect to $\tilde r$. The functions
$\tilde a$, $\tilde b$, $\tilde c$, $\tilde h$ and $\tilde k$ come from
the functions $a(r)$, $b(r)$, $c(r)$, $h(r)$ and $k(r)$ in (3.10)
respectively by variables
replacement as ${\tilde x}_0 =m_f^{} x_0$ for $x_0$, ${\tilde r} =m_f^{} r$
for $r$ and ${\tilde \rho} =m_f^{} \rho$ for $\rho$.
There are two parameters ${\tilde x}_0$ and $\tilde \rho$ in these equations.
In the following calculation, we will limit $\rho$ to take the values in
formula (3.14) which makes the barrier height of instanton minimum.
So there is only one parameter $\tilde x_0$ in the equations (4.5).
\par
To solve the eigenvalue equations (4.5) for the heavy fermions in the
background field of instanton, it is required certain boundary conditions
for the fermion wavefunctions. Wavefunctions $G_L$ and $G_R$ are finite
and $F_L$ and $F_R$ are zero at $\tilde r =0$ and that all wavefunctions
$G_L$, $G_R$, $F_L$ and $F_R$ tend to zero in the limit
$\tilde r \to \infty$.

\vskip 3pc

\centerline{\bf V. ANOMALOUS DECAY OF HEAVY FERMION}
\vskip 2pc
Let us consider the system of a fermion doublet in the background field
of electroweak instanton.
As the Euclidean time $x_0$ changes from $-\infty$ to
$+\infty$, the one-instanton field evolves from one vacuum to another
topologically distinct vacuum. If the system starts with the filled Dirac
sea plus one real fermion, then the final state has the filled Dirac sea
and no real fermion, so that the fermion number is not conserved.
For the fermion with small mass, the process of fermion number
non-conservation is exponentially suppressed. However, if the fermion
is sufficiently heavy, instead of tunneling from one vacuum to
another topologically distinct vacuum, the fermion might freely move.
So one can expect that rapid anomalous decay of the heavy fermion
occur classically.
\par
To investigate the possibility of unsuppressed fermion number violation
in heavy fermion anomalous decay,
we solve numerically the fermion eigenvalue equations
(4.5) under the boundary conditions for bound state in the background
field of instanton with minimum barrier.
At first, we take the parameter $\alpha=1$ ($M_H=M_W$) and the size of
minimum barrier of instanton $\rho_0$ is given by (3.14) with $\rho_0=2.98/v$.
Giving a heavy fermion mass and a values for parameter $\tilde x_0$
, which stand for the Euclidean time $x_0$ for fixing fermion mass $m_f^{}$,
we can numerically solve the equations by computer and obtain the eigenvalue
$\tilde \epsilon$ and eigenfunctions $F_L$, $G_L$, $F_R$ and $G_R$.
\par
In the beginning, let us calculate the process of heavy fermion level
crossing.
There are only $\tilde\epsilon =0$ solutions in the case of $x_0 =0$
at which the Chern-Simons number $N_{CS} =1/2$.
These normalizable eigenstates with zero eigenvalue are the
zero-mode of fermion in the background fields of instanton that have
discussed by authors[3,10]. To see that,
we make a gauge transformation for the left-hand zero mode eigenfunctions
$$
       q'_L (x)= V (x)q_L (x) ,  \eqno  (5.1)
$$
by using the expression (3.6) for $V(x)$ and (4.3) for $q_L$. We obtain
$$
\eqalign{  G'_L &=G_L\cos f +F_L \sin f ,  \cr
           F'_L &=-G_L\sin f +F_L \cos f .
}    \eqno   (5.2)
$$
Comparing the functions $G'_L$ and $F'_L$ with the corresponding analytical
solutions given by Rubakov[3] for the fermion zero mode, one found that
it coincide with each others respectively.
\par
We solve the fermion eigenvalue equations (4.5) for non-zero values of
parameter $x_0$ which can vary from $-\infty$ to $+\infty$.
Since the configurations of instanton are
symmetric about $x_0 =0$ at which the zero mode appears, the fermion
eigenvalue $\tilde \epsilon$ should be antisymmetric with respect to the
$x_0 =0$ configuration.
\par
In Fig. 2 we represent the fermion eigenvalues $\tilde \epsilon$ for
dependence of the Chern-Simons number $N_{CS}$ as the Yukawa coupling
constant of heavy fermion $f_q=$65 in the background field of
minimum barrier with $\rho_0=2.98/v$ for $M_H=M_W$.
Fig. 2 shows the whole process of
fermion level crossing that is from positive-energy continuous state
to negative-energy continuous state.
Since $N_{CS}$ is a function of $x_0$, Fig. 2 also represents the
eigenvalue $\tilde \epsilon$ dependence of
$x_0$ as $N_{CS}$ varying from 0 to 1 correspond to the $x_0$
varying from $-\infty$ to $+\infty$ and the gauge field configurations
changing from one vacuum to another vacuum.
\par
During the process of level crossing of heavy fermion, let us consider
the total energy change of this system.
In the classical approximation, the total energy is the sum of
bosonic energy $E_b$ and energy of the fermion occupying
a energy level $E_f$
$$
         E_t(x_0)=E_b(x_0)+E_f(x_0),        \eqno (5.3)
$$
where $E_b$ is given by (3.12) and $E_f=m_f \tilde\epsilon$. If the gauge
and Higgs fields take their vacuum values, $E_f$ is the mass of the fermion.
During the process of fermion level crossing, we can evaluate the total
energy $E_t$ at each value of parameter $x_0$.
The dependence of the total energy $E_t$
on the Chern-Simons number $N_{CS}$ along the minimum barrier of instanton
with $\rho_0=2.98/v$ (for $M_H=M_W$) is shown in
Fig. 3 for several values of Yukawa coupling constant $f_q$ of heavy fermion.
Fig. 3 describes the behaviores of total energy $E_t$ in heavy fermion level
crossing: at the beginning the total energy tends to mass of fermion;
at the top of the barrier the fermion level is at the zero mode and
the total energy equal to the height of the barrier; at the end the total
energy tends to the minus mass of the fermion in the Dirac sea.
We found that the behaviores of total energy $E_t$ depending on
$N_{CS}$ can be separated into three types:
\par
At $f_q \leq f_q^0=25.0$ ( $m_f^0=4.35$ TeV ),
the state corresponding to the fermion is separated
from the topologically non-trivial vacuum state by the barrier
of electroweak instanton. So the fermionic state is metastable,
its decay proceeds via tunneling
and the decay amplitude is exponentially suppressed.
\par
At $f_q^0 < f_q < f_q^{\rm cr}=52.8$ ( $m_f^{\rm cr}=9.2$ TeV ),
the total classical energy has a local minimum between the mass of fermion
and the barrier height of instanton. Due to the barrier also existes,
the anomalous decay only is by tunneling and the decay amplitude is small.
\par
At $f_q \geq f_q^{\rm cr}$,
the classical total energy monotonically decreases along the path of the
barrier of instanton.
Hence the evolution of the system along this path is not classically
forbidden. The initial state, corresponding to $x_0$ at $-\infty$, describes
the heavy fermion and trivial vacuum of the bosonic fields. As the parameter
$x_0=0$, the bosonic fields at the top of minimum barrier with $N_{CS}=1/2$
, while the fermion level crosses zero and entry into the
negative energy ranges. The final state, corresponding to $x_0$ at
$+\infty$, describes the topologically non-trivial vacuum, containing
no real fermions.
\par
So we found that, if the mass of the heavy fermion exceeds the critical
value $m_f^{\rm cr}=9.2$ TeV in the case of $M_H=M_W$, its anomalous
decay might proceed without tunneling  and its lifetime might be small.
Within the classical approximation, we can conclude that the fermion number
violation in the heavy fermion decay is not exponentially suppressed for
sufficiently large $m_f$. This value of the critical mass of the
heavy fermion is agreement with the naive estimate of ref. 3 and
the variational calculation of ref. 9, but our values is exactly
numerical results.
\par
In the above calculation, we fix the mass of Higgs boson at $M_H=M_W$
(for $\alpha=1$)
with the size of instanton at $\rho_0=2.98/v$ for minimum barrier height.
Now we change the mass of Higgs boson for calculation.
The critical fermion mass $m_f^{\rm cr}$ depending on the mass of Higgs boson
is shown in the Fig. 4. We found that the value of the critical mass
$m_f^{\rm cr}$ for unsuppressed fermion number violation tends to a constant
for small values of Higgs mass and goes to large as the Higgs mass
becoming large.
In the range of weak Higgs self-coupling, the critical mass of the heavy
fermion is small than 15 TeV. If the fourth generation fermion exists,
due to the large mass hierarchy between generations, the heavy fourth quarks
might exist the unsuppressed anomalous decay.

\vskip 3pc

\centerline{\bf VI. CONCLUSIONS}
\vskip 2pc
The detailed behavior of the total classical energy of heavy fermion
in the background field of the electroweak instanton have been
demonstrated by numerical calculation. The anomalous decay path
of heavy fermion are also presented. Within the classical approximation,
it is shown that the standard electroweak model with a fourth generation
fermion doublet might appear the unsuppressed fermion number violation
due to the heavy fermion anomalous decay.
The dependence of the decay rate on the mass of the heavy fermion is
extremely sharp: for masses below the critical value $m_f^{cr}$
($m_f^{cr}=9.2$ TeV for $M_H=M_W$),
the decay process occurs by tunneling and the decay probability is
exponentially suppressed, while for masses exceeding the critical value
the decay proceeds without tunneling, and the lifetime is small.
The dependence of critical mass $m_f^{cr}$ on the Higgs mass is
also presented.
\par
Unfortunately, within the classical approximatiom,
the arguments presented in this paper are far from final conclusion.
Because we have neglected the radiative corrections due to
the boson loops and the contribution of the Dirac sea throughout the
whole discussion. Even more to large Yukawa coupling
(i.e. for large fermion mass),
at present, one cannot analyze this strong coupling theories in a reliable
way, so we cannot justify this approximation.
Nevertheless we think that our results are the first step to study
the problem of fermion number violation in the heavy fermion decay.
\par
If the fourth generation fermion exist, it might be very heavy.
The heavy fermion in the standard model will violate the vacuum stability
due to the radiative corrections. So we must improve the standard model
with new physics in the TeV region, for example,
introducing extra bosonic fields with almost degenerate with fermions;
supersymmetry, and so on. On the other hand,
the fermion number violation mechanism
in the standard electroweak theory is so fundamental and natural,
so we hope that fermion number violation in the heavy fermion
anomalous decay can appear rapidly even in the improved standard model.

\vskip 2pc
\centerline{\bf ACKNOWLEDGMENTS}
\vskip 1pc
The author is indebted to Dr. D. H. Zhang for help in drafting on computer.
I am very grateful to Professor T. Huang for warm hospitality and
Professor J. M. Wu for kindly help.

\vfill\eject

\centerline{\bf REFERENCES }
\vskip 3pc

\item{1} CDF Collaboration, F.Abe $et\quad al.$, \prl {\bf 74},2626(1995).
\item{2} Particle Data Group, \pr {\bf D50}, 1433(1994).
\item{3} V. A. Rubakov, \np {\bf B256}, 509(1985).
\item{4} G. 't Hooft, \prl {\bf 37}, 8(1976); \pr {\bf D14}, 3432(1976).


\item{5} R. Jackiw and C. Rebbi, \prl {\bf 37}, 172(1976);
\item{ } C. G. Callan, R. F. Dashen and D. J. Gross, \pl {\bf B63}, 334(1976).


\item{6} V. A. Matveev, V. A. Rubakov, A. N. Tavkhelidze, and
         M. E. Shaposhnikov,
\item{ } Usp. Fiz. Nauk {\bf 156}, 253(1988).
\item{7} V.A. Kuzmin, V.A. Rubakov and M.E. Shaposhnikov, \pl
         {\bf B155}, 36(1985);
\item{ } P. Arnold and L. McLerran, \pr {\bf D36}, 581(1987);
\item{ } A. Ringwald, \pl {\bf B201}, 510(1988).
\item{8} A. Ringwald, \np {\bf B330}, 1(1990);
\item{  } O. Espinosa, \np {\bf B343}, 310(1990).
\item{9} G. G. Petriashvili, \np {\bf 38}, 468(1992),
         and references therein.


\item{10} C. G. Callan, R. F. Dashen and D. J. Cross, \pr {\bf D17},
          2717(1978);
\item{  } J. Kiskis, \pr {\bf D18}, 3690(1978).
\item{  } N. H. Christ, \pr {\bf D21}, 1591(1980);
\item{  } J. Ambjorn, J. Greensite and C. Peterson, \np {\bf B221}, 381(1983).
\item{  } N. V. Krasnikov, V. A. Matveev, V. A. Rubakov, A. N. Tavkhelidze
\item{  } and V. F. Tokarev, Teor. Mat. Fiz. {\bf 45}, 313(1980).
\item{11} K. Yang, \pr {\bf D49}, 5491(1994).


\item{12} N. V. Krasnikov, Yad. Fiz. {\bf 28}, 549(1978);
\item{  } P. C. Huang, \prl {\bf 42}, 873(1979);
\item{  } H. D. Politzer and S. Wolfram, \pl {\bf B82}, 442(1979);
                                             {\bf B83}, 42(1979).
\item{13} N. S. Manton, \pr {\bf D28}, 2019(1983).
\item{  } F. R. Klinkhamer and N. S. Manton, \pr {\bf D30}, 2212(1984).
\item{14} A. A. Belavin, A. M. Polyakov, A. S. Schwarz and Yu. S. Tyupkin,
\item{  } \pl {\bf B58}, 85(1975).
\item{15} I. Affleck, \np {\bf B191}, 429(1981).


\vfill\eject


\centerline{\bf FIGURE CAPTIONS}
\vskip 2pc

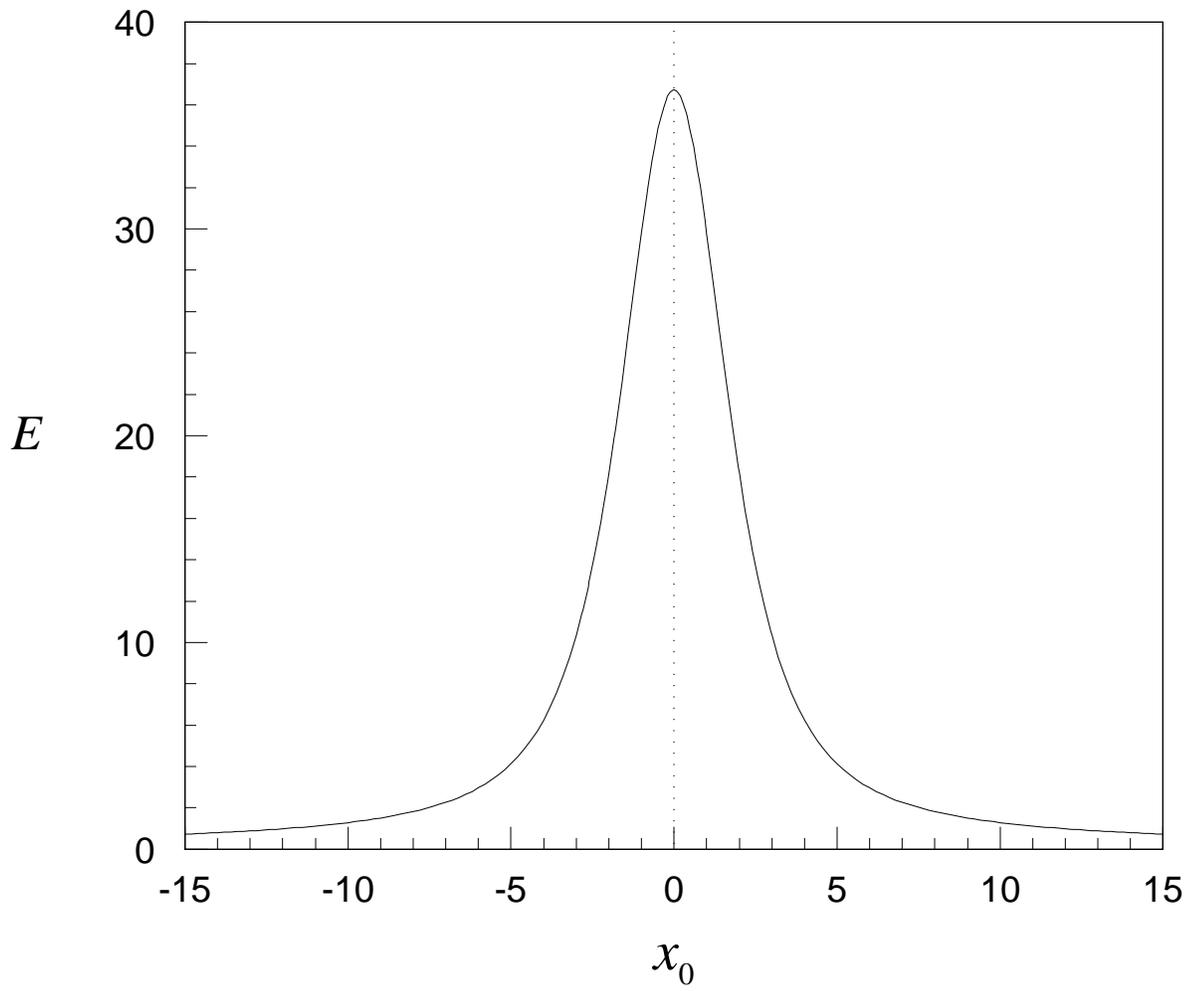
\item{\bf FIG. 1} The minimum barrier of instanton with the size
                  $\rho_0=2.98/v$: the energy (unit in $v$) of
                  gauge and Higgs fields is shown as a function of
                  Euclidean time $x_0$ (unit in $1/v$).
                  Here, we have taken $M_H=M_W$ ($\alpha=1$).
\vskip 1pc
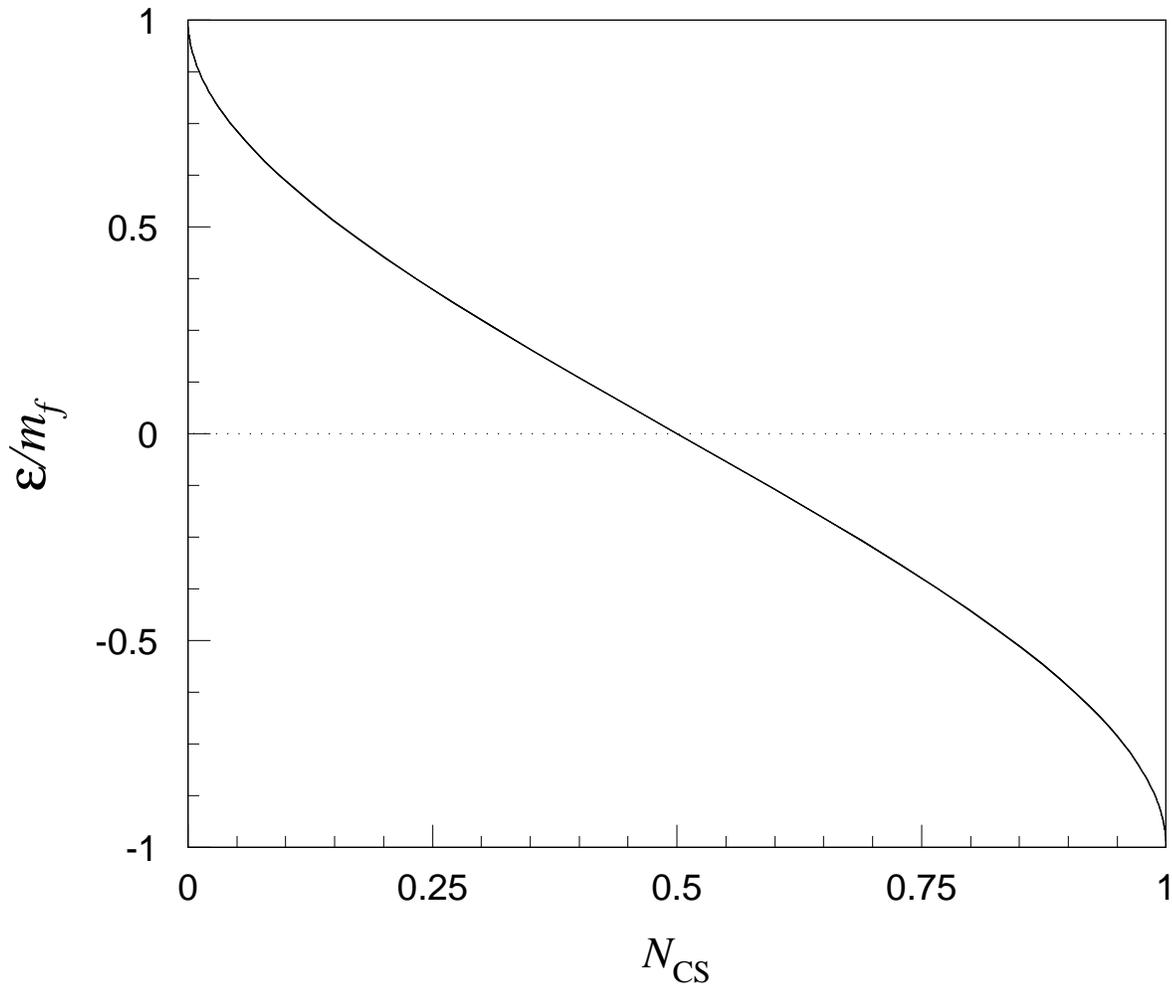
\item{\bf FIG. 2} In the minimum barrier of instanton with $\rho_0=2.98/v$,
                  the normalized fermion eigenvalue
                  $\tilde\epsilon=\epsilon/m_f$ is shown as a function
                  of the Chern-Simons number $N_{CS}$ which changes from zero
                  to one for the Yukawa coupling constant of heavy fermion
                  $f_q=65$.
\vskip 1pc
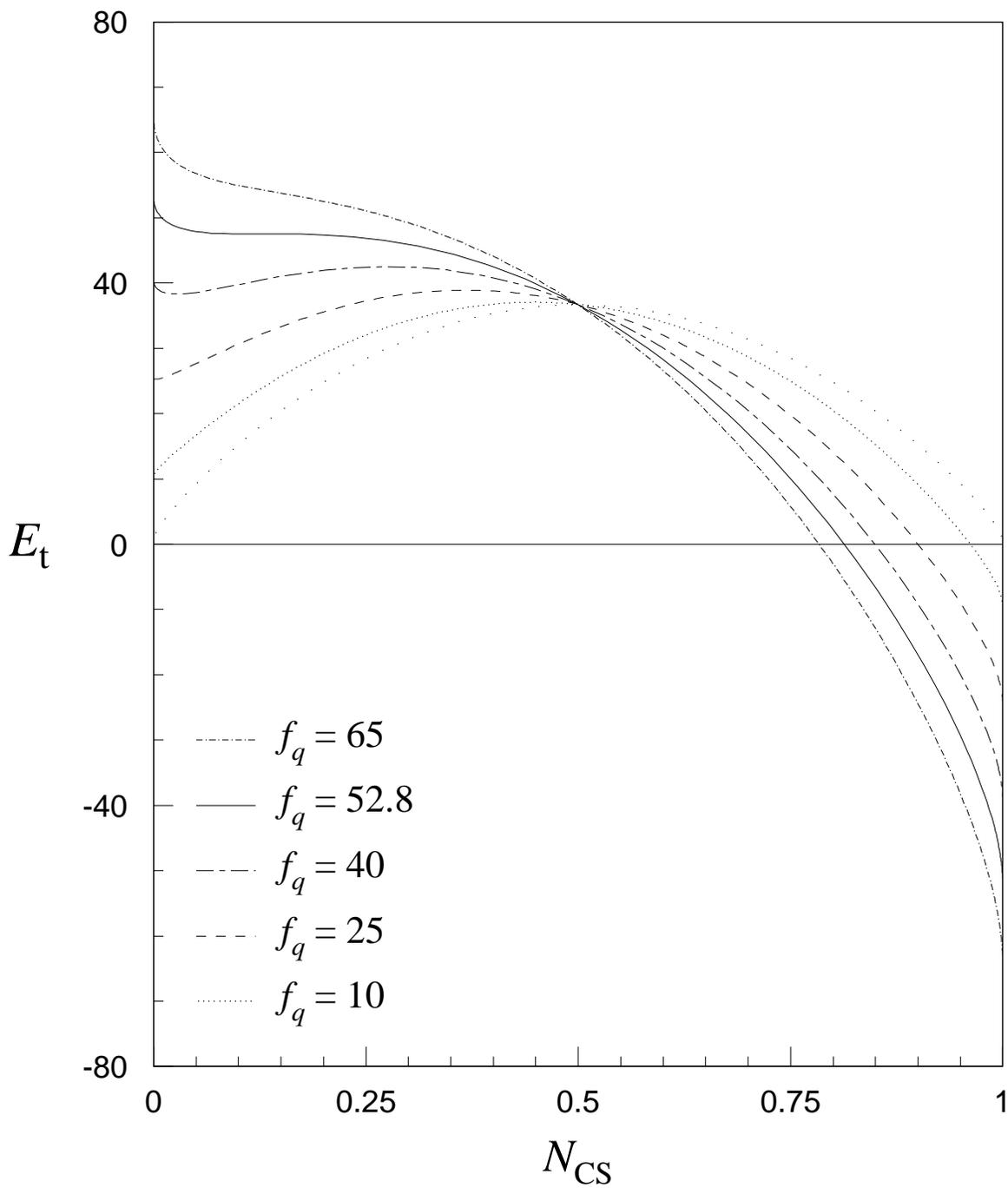
\item{\bf FIG. 3} The total energy (unit in $v$) as a function of the
                  Chern-Simons number $N_{CS}$ along the minimum barrier
                  of instanton in the case of $M_H=M_W$ ($\alpha=1$)
                  for Yukawa coupling constant of fermions taking
                  $f_q$=65, 52.8, 40, 25, 10.
                  We obtained $f_q^{\rm cr}=52.8$ (in solid) and
                  $f_q^0=25.0$ (in dashed).
                  The minimum barrier of instanton with $\rho_0=2.98/v$
                  is also shown (in sparsely dotted).
\vskip 1pc
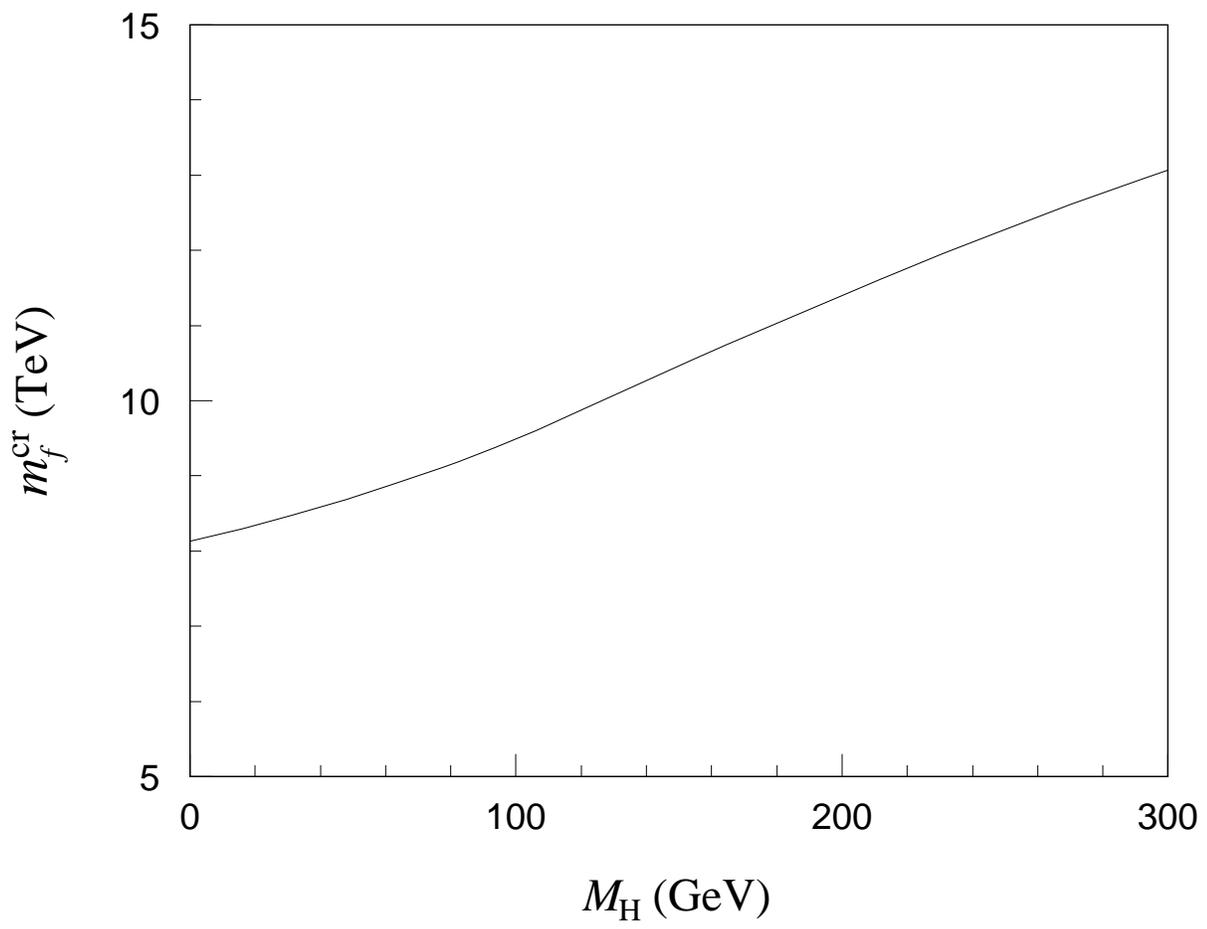
\item{\bf FIG. 4} The critical fermion masses for unsuppressed fermion
                  number violation depend on the Higgs mass.

\end